# Understanding Polarized Foreground from Dust: Towards Reliable Measurements of CMB Polarization

A Science White Paper for the:
**Cosmology and Fundamental Physics (GCT) Science Frontiers Panel**
of the Astro2010 Decadal Survey Committee

Lead Author:

Alex Lazarian
Astronomy Department
475 Charter St.
University of Wisconsin
Madison
(6o8) 262 – 1715 (ph)
lazarian@astro.wisc.edu

Dan Clemens
Astronomy Department
725 Commonwealth Ave
Boston University
Boston, MA 02215
(617) 353 – 6140 (ph)
clemens@bu.edu

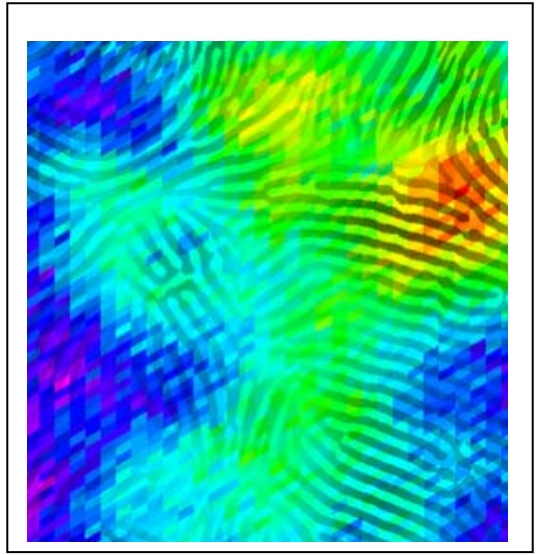

Figure at right: Simulated CMB temperature fluctuations and the direction of the CMB partial polarization. The direction of the stripes indicates the direction of polarization. Darker stripes indicate stronger polarization (technique by Cabral & Leedom 93).

Contributors and Signatories:

| | |
|---|---|
| Andy Adamson | UKIRT, JAC, Hilo |
| B.-G. Andersson | SOFIA, NASA Ames |
| David Axon | Rochester Institute of Technology |
| James De Buizer | SOFIA, NASA Ames |
| Alberto Cellino | Osservatorio Astronomico di Torino, Italy |
| Dean C. Hines | Space Science Institute, Corrales, NM |
| Jennifer L. Hoffman | University of Denver |
| Terry Jay Jones | University of Minnesota |
| Antonio Mario Magalhaes | University of Sao Paulo, Brazil |
| Joseph Masiero | University of Hawaii |
| Chris Packham | University of Florida |
| Marshall Perrin | UCLA |
| Claudia Vilega Rodrigues | Inst. Nac. De Pesquisas Espaciais, Brazil |
| Hiroko Shinnaga | CalTech |
| William Sparks | STScI |
| John Vaillancourt | CalTech |
| Doug Whittet | RPI |

## *CMB Polarization and Polarization from Aligned Galactic Dust*

Measurements of the Cosmic Microwave Background (CMB) fluctuations and the interpretation of those fluctuations have played a leading role in establishing the present day model of the universe we live in. This modern day scientific revolution was only possible through careful accounting of the various sources of foreground emission and absorption that could have hidden the weak fluctuation signal from the distant past.

Polarization of the CMB provides information about the universe that is not contained in the temperature data alone. In particular, it offers a unique way to trace the primordial perturbations of a tensorial nature (e.g., cosmological gravitational waves), and allows breaking important degeneracies that plague measurement of cosmological parameters with intensity alone (see Zaldarriaga et al. 1997).

The polarization of CMB arising from the curl-free polarized E-modes, which required a factor of 100 more sensitivity, has been also reported (Page et al. 2007). The next challenge is the measurement of the signatures of primordial gravitational waves, which are predicted to leave a distinct divergence-free, or B-mode, pattern in the large-scale CMB polarization anisotropy. The expected signal is already constrained to be an order of magnitude smaller than the observed E-mode signal. This level of signal is potentially detectable by the Planck satellite and ground and balloon-based experiments, *provided the polarized CMB foregrounds are carefully taken into account*. Once B-modes are detected, the research focus will shift to their detailed characterization, which would require even higher accuracy of the foreground removal process.

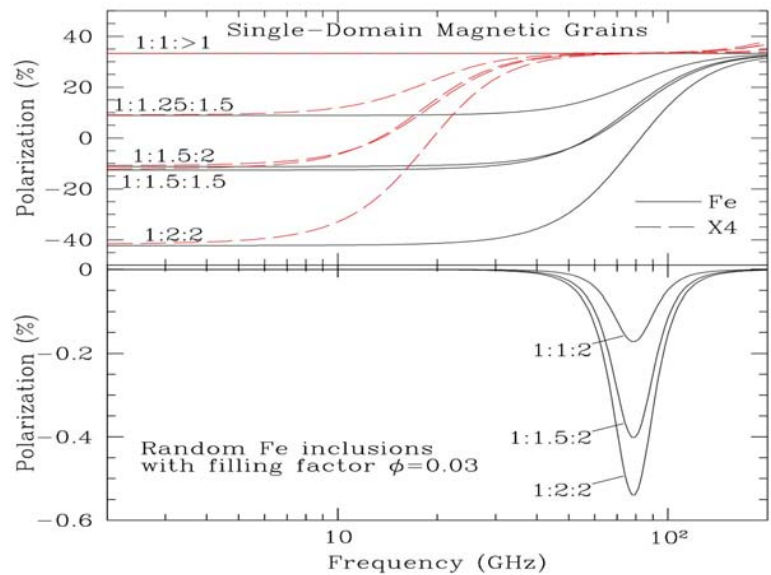

Figure1. Polarized magneto-dipole emission from magnetic aligned grains. Upper panel: Polarization of thermal emission from perfectly aligned single domain grains of metallic Fe (solid lines) or hypothetical magnetic material that can account for the anomalous emission. Lower panel: Polarization from perfectly aligned grains with Fe inclusions (filling factor is 0.03). Grains are ellipsoidal and the results are shown for various axial ratios. From Draine & Lazarian

Among different sources of polarized foregrounds, the one arising from aligned interstellar dust is believed to be the most difficult to correct (see Lazarian & Finkbeiner 2003, Dunkley et al. 2008). As the degree of dust alignment is known to depend both on grain size and grain composition, this produces a frequency dependence to the dust polarization. Moreover, in addition to the "classical" power-law distribution of large grains (see Mathis, Rumpl & Nordsieck 1977), dust has a population of tiny grains (Leger & Puget 1984), often called PAHs. These tiny grains have been identified by Draine & Lazarian (1998) as a source of anomalous emission (Kogut et al. 1996), which raises important questions about the possible polarization from this emission mode. In addition, aligned grains with inclusions of material with strong magnetic response, are predicted to emit polarized magneto-dipole emission with their own complex frequency dependent polarization. The polarization may also be very high, exceeding 40% (see Figure 1; Draine & Lazarian 1999).

In view of the vital importance of separating the CMB polarized signals from the various polarized foregrounds, coordinated efforts of observers and theorists are needed to establish reliable constraints on the fraction of microwave polarization arising from the foreground dust. Such efforts need only small resources compared to those of a single NASA mission. Yet, the improvement in understanding of the properties of the polarized dust foreground can substantially improve the quality of modeling and lead to precision removal of the foreground from the critical data from many different CMB experiments.

The time is ripe for a breakthrough in this area. First, dramatic improvements in the quantity and quality of measured ground-based interstellar polarimetry has been achieved from the UV though the submm. Second, there has been a substantial improvement of our understanding of grain alignment (see Lazarian 2007). For the first time, quantitative predictions of grain alignment are possible and are testable by observations (see Whittet et al. 2007). Moreover, progress has been achieved in understanding the nature of foreground polarization fluctuations (see Fosalba et al. 2002), for instance, through relating their spatial spectrum to the Kolmogorov-type spectrum of the underlying interstellar turbulence (Cho & Lazarian 2002, 2008).

These recent successes motivate a strong program of research aimed at quantitative description of the polarized radiation from dust. The program must combine better high sensitivity observations of the polarized emission from dust as well as further improved modeling of polarized microwave emission from dust. The latter can be achieved by dedicated testing of the theoretical predictions available.

### *Example Polarimetry Science Areas for the next decade*

Within the area of CMB polarization studies, polarimetry from ground-based observatories at different wavelengths needs to be combined with theoretical studies, numerical modeling and laboratory studies to be able to address the following **key questions**: *What is the wavelength dependence of the foreground polarized microwave emission arising from aligned Galactic and Zodiacal dust? What are the statistical properties of polarized emission from Galactic dust? How important is the polarized radiation arising from the magneto-dipole emission of strongly magnetic dust grains? What is the degree of polarization of the anomalous microwave emission?* In the following, example studies aimed at answering some of these questions are presented.

- Observational studies of the wavelength dependency of polarization.

    Unpolarized emission has been well characterized using data from IRAS, allowing this emission to be removed from the CMB signal (e.g., Kogut et al. 2007). However, no all-sky polarization measurements exist to be used to remove Galactic dust polarization from CMB observations.  In fact, extrapolation of percentage polarization from one frequency to another with high accuracy requires sophisticated modeling that employs grain alignment theory. A direct extrapolation of the multi-temperature dust model to microwave frequencies suggests a nearly constant polarization spectrum, but existing FIR polarization measurements are limited to dense clouds.  These measurements do show rather complex dependences of polarization with the wavelength, arising from grains at different temperatures aligning differently.

    Future FIR through mm wavelength observations (especially SOFIA at 50 - 215 $\mu$m near the dust emission peak, and *Planck* at 80 - 850 GHz) will target the diffuse regions of the ISM, providing valuable input for accurate modeling of microwave polarization from dust. The synergy of the ground-based and SOFIA observations with the Planck mission, which uses a wider frequency range (80 - 850 GHz) should provide polarization radiation templates and important constraints for grain alignment theory.

- Constructing models of polarized microwave dust emission

    If the composition of grains is known, their polarization depends on the grain shape and grain alignment. The latter is the more complex effect. One may expect variations of alignment depending on grain environment, which can introduce substantial variations of the polarization with frequency. Therefore, it is essential to provide extensive testing of grain alignment. This can be performed at many different wavelengths, including optical and near infrared.

    Modern grain alignment theory is based on radiative torques. The efficiency of those decreases as the grain size gets small compared to the radiation wavelength. As a result, small grains are not expected to align well; the cut-off size of aligned grains depends on the intensity, as well as the spectrum and degree of anisotropy of the radiation illuminating the dust grain. The measurements of polarization (Whittet et al. 2008; Hildebrand et al. 2000) agree well with simple modeling. However, more detailed, multi-frequency, measurements of polarization are required.

    Theoretical modeling also should be improved. For instance, an additional dependence, not yet included in any modeling of polarization to date, is on the angle between the direction of anisotropic flux of photons and the magnetic field (Lazarian & Hoang 2007). Understanding how the aligned grains size distribution depends on environment is one of the key goals for quantitative modeling of dust polarization. The correct degree of polarization is another input parameter needing theoretical exploration. We discuss this below.

- Determining Degree of Alignment and Presence of Strongly Magnetic Grains

The theory predicts the degree of paramagnetic grain alignment as a function of the angle between the direction between the anisotropic flux of photons that align grain and the magnetic field (Lazarian & Hoang 2007). However, if grains have strongly magnetic inclusions (see above) the degree of alignment is 100% and does not change with the aforementioned angle (Lazarian & Hoang 2008). A dedicated program of observations and simulations can distinguish the two cases and provide constraints on the presence of the strongly magnetic grains or strongly magnetic inclusions in interstellar grains.

If grains are strongly magnetic they will exhibit 100% alignment and this will increase the modeled polarization values for frequencies larger than 100 Ghz. However, this will mean that in the range of 20-100~Ghz, additional polarized magneto-dipole emission is expected[1].

Recent observational probing of grain alignment (Andersson & Potter 2007) shows that observational testing of the predictions is feasible in the optical and infrared ranges. A dedicated program of multi-frequency high-resolution observations, combined with modeling of the 3D distribution of the radiation sources, i.e. stars, is required to test grain alignment theory, solve this long-standing (60 years old!) puzzle, and provide an accurate model of dust polarization.

- Finding the polarization of spinning dust emission

    The existing theory of PAHs alignment predicts marginal degrees of alignment for grains rotating at higher than 30 Ghz frequencies (Lazarian & Draine 2000). Testing can be performed in the microwave and in the IR, as the smallest grains will emit there. This requires new instruments and new polarimetric studies. Laboratory studies of the paramagnetic response of the PAHs are also needed in order to test the effect of resonance paramagnetic relaxation predicted in Lazarian & Draine (2000).

- Role of Laboratory studies

    Not only the paramagnetic response of PAHs requires additional studies. The predictions of the magneto-dipole emission in Draine & Lazarian (1999) are based on data on magnetic response at microwave frequencies that they could get from the literature. The necessity of more precise predictions calls for dedicated laboratory studies of microwave response of candidate magnetic materials.

- Determining the statistics of interstellar dust polarization

    To separate the CMB polarization signal and polarized foreground contribution arising from aligned dust, it is important to know the expected statistical properties of the foreground polarized emission. Those can be evaluated from the statistics of the starlight polarization (Fosalba et al. 2002), provided that the difference in the sampling of volume with aligned grains by stars and the microwave emission is taken into account. Multi-frequency observations of the polarized emission provide another way of

---

[1] Even paramagnetic grains are expected to emit polarized magneto-dipole emission, although at a lower rate (Draine & Lazarian 1999). This emission in the range below 100 Ghz will become important as CMB polarization measurements get more precise.

separating the CMB and foreground polarization. Studies of the statistics of the starlight polarization in Cho & Lazarian (2002, 2008) showed that the underlying spectrum of magnetic fluctuations, those that have the largest influence on the variations of the polarization degree in diffuse media, is consistent with the Kolmogorov-type spectrum. The deviation in the value of the spectral index of the polarization fluctuations obtained in Fosalba et al. (2004) from the standard 5/3 Kolmogorov value was shown to arise from the volume sampling by stars, the majority of which are close to the observer. Better star polarization surveys, e.g. the one by Clemens et al. (2009) as well as far-infrared polarimetry studies of individual molecular clouds should allow for accurate testing of the hypothesis of the Kolmogorov spectrum of dust polarization. The knowledge of the underlying spatial polarization spectrum of emission arising from dust should help to separating its contribution from that of B-modes of CMB.

## Broad Impact: Synergy of studying CMB polarization and interstellar magnetic fields

While measuring the polarization of CMB B-modes is a challenging problem along the path toward understanding the early universe, the detailed mapping of magnetic fields is another challenging problem, but one needed for understanding the role of the field in many astrophysical processes, including the propagation of cosmic rays, star formation, and heating of the interstellar gas. Tracing magnetic fields with aligned grains provides a unique window for revealing magnetic fields in diffuse interstellar gas and molecular clouds. Resolving the remaining uncertainties related to grain alignment as well as obtaining high-density maps of magnetic fields in dense gas, and, importantly, in high latitude diffuse gas, will shed light on the many astrophysically-important effects of magnetic fields.

## Context: Polarimetry as a cross-cutting enterprise

Polarimetry is practiced across the full range of accessible wavelengths, from long wavelength radio through gamma rays, to provide windows into phenomena not open to photometry, spectroscopy, or their time-resolved variants. At some wavelengths, the U.S. leads the world in polarimetric capabilities and investigations, including ground-based radio, through the VLA and VLBA. At other wavelengths, the U.S. is currently competitive: in submm the CSO and the JCMT have historically pursued similar science problems.

In ground-based O/IR, the situation is considerably worse, with no optical or NIR polarimeters available on Gemini (Michelle is MIR, only) or any NOAO-accessed 4 m telescope, as the table below shows. Over the past decade and more, Canadian and European astronomers have enjoyed unique access to state-of-the-art polarimeters and have used this access to vault far past the U.S. in many science areas.

Photometry, spectroscopy, and polarimetry together comprise the basic toolbox astronomers use to discover the nature of the universe. Polarimetry established the Unified Model of AGN and continues to yield unique and powerful insight into complex phenomena. Polarimetry reveals the elusive magnetic field in the Milky Way and external galaxies, allows mapping of features of unresolved stars and supernovae, uncovers nearby exoplanets and faint circumstellar disks, and probes acoustic oscillations of the early universe.

| Telescope | Aperture | Instrument | Waveband | Polar. Mode | U.S. Access ? |
|---|---|---|---|---|---|
| IRSF (SAAO) | 1.4m | SIRPOL | NIR | Imaging | No |
| Perkins (Lowell) | 1.8m | Mimir, PRISM | NIR, Optical | Imaging | Private |
| HST | 2.4m | WFPC2, ACS, NICMOS | Optical, Optical, NIR | Imaging | Yes |
| Nordic Optical | 2.5m | TURPOL | Optical | Photopol | No |
| MMT | 6.5m | MMTPOL | NIR | Imaging | Private |
| LBT | 2x8.4m | PEPSI | Optical | Spectropol | Private |
| Gemini | 8m | Michelle | MIR | Imaging | Yes |
| Keck | 10m | LRIS | Optical | Spectropol | Private |
| GTC | 10m | CanariCam | MIR | Imaging | No |

In space, NICMOS, ACS, and WFPC-2[2] on HST have permitted imaging polarimetry down to 0.1% precision, and may represent the most general purpose O/IR access for U.S. astronomers. Neither the Spitzer Space Telescope nor JWST provides, or will provide, any polarimetric capability.

The dwindling U.S. access to this crucial third leg of the light analysis tripod has also become self-fulfilling, as students receive little exposure to polarimetric techniques and scientific advances as the number of practitioners able to teach students declines.

## *Final Thought*

The U.S. astronomical community has lost opportunities to advance key science areas as a result of down-selects of instrument capabilities or lack of will to commission polarimetric modes on instruments. The investment is minor, the expertise is available in the community, and the rewards are tangible. We are excited by the recent momentum favoring polarimetric studies and capabilities and believe the upcoming decade will see the various polarimetric techniques together become a strong, necessary component of astronomers' light analysis toolbox. In terms of achieving progress in modeling polarized microwave emission, we argued above that dedicated efforts involving multi-frequency polarimetry, including optical and infrared polarimetry are required.

*Bibliography and References*
Andersson, B.-G., & Potter, S. B. 2007, ApJ, 665, 369
Cho, J., & Lazarian, A. 2002a, ApJ, 575, L63
Cho, J., & Lazarian, A. 2008, arXiv:0812.2023
Clemens, et al. 2009, BAAS, 213, 325.04
Draine, B.T. & Lazarian A. 1998a, ApJ, 494, L19
Draine, B.T. & Lazarian 1999, ApJ, 512, 740-754
Dunkley et al 08: http://arxiv.org/abs/0811.3915
Fosalba, P., Lazarian, A., Prunet, S., & Tauber, J. A. 2002, ApJ, 564, 762
Fraisse et al 2008: http://arxiv.org/abs/0811.3920
Hildebrand et al. 2000, PASP, 112, 1215

---

[2] NICMOS and much of ACS are currently off-line until Servicing Mission 4 (SM4); WFPC-3 will replace WFPC-2 but will have no polarimetric capability.

Kogut, A., Banday, A. J., Bennett, C. L., Gorski, K. M., Hinshaw, G., & Reach, W.~T. 1996, ApJ, 460, 1
Kogut, A. et~al. 2007, ApJ, 665, 355
Lazarian, A., & Draine, B.T. 2000, ApJ, 536, L15
Lazarian, A. & Finkbeiner, D. 2003, NewAR, 47, issue 11-12, 1107
Lazarian, A. & Hoang, T. 2007, MNRAS, 378, 910
Lazarian, A., & Hoang, T. 2008, ApJL, 676, L25
L'eger, A., & Puget, J.L. 1984, ApJL, 278, L19
Mathis, J.S., Rumpl, W., & Nordsieck, K.H. 1977, ApJ, 217, 425
Page et al. 2007, ApJ Supplement, 170, 335
Whittet, D. C. B., Hough, J. H., Lazarian, A., & Hoang, T. 2008, ApJ, 674, 304
Zaldarriaga, M., Spergel, D.N., & Seljak, U. 1997, ApJ, 488, 1